\documentclass[aps,eqsecnum,preprint,preprintnumbers,12pt,amsfonts]{revtex4}

\usepackage{epsfig,psfrag}
\usepackage{latexsym}

\def\beq{\begin{eqnarray}}
\def\eeq{\end{eqnarray}}
\def\bea{\begin{eqnarray}}
\def\eea{\end{eqnarray}}

\newcommand\eqn[1]{(\ref{#1})}      
\newcommand{\nn}{\nonumber}
\newcommand{\reals}{\mbox{${\rm I\!R }$}}

\newcommand{\intgs}{\mbox{${\rm Z\!\!Z }$}}
\newcommand{\ep}{\epsilon}
\newcommand{\kh}{\hat{\mathbf{k}}}

\newcommand{\intl}{\int\limits}
\newcommand{\sumne}{\sum_{n=1}^\infty}
\newcommand{\sumnm}{\sum_{n=-\infty}^\infty}
\newcommand{\sumi}{\sum_{i=1}^\infty}
\newcommand{\sumip}{{\sum_{i=1}^\infty}'}
\newcommand{\lis}{\lambda_i^2}
\newcommand{\sumtop}{{\sum_{n_1,...,n_d=-\infty}^\infty}\hspace{-0.5cm}'\hspace{.50cm}}
\newcommand{\spt}{n_1^2+...+n_d^2}
\newcommand{\sumdp}{{\sum_{n_1,...,n_d=-\infty}^\infty}^{\hspace{-.5cm}\prime\hspace{.5cm}}}

\usepackage{color}

\begin{document}

\title{Kaluza--Klein models as pistons}
\author{Klaus Kirsten}\email{klaus_kirsten@baylor.edu}
\affiliation{Department of Mathematics, Baylor University, Waco,
TX 76798-7328 USA}
\author{S. A. Fulling}\email{fulling@math.tamu.edu}
\affiliation{Departments of Mathematics and Physics, 
Texas A\&M University, College Station, TX, 77843-3368 USA}
\date{\today}

\begin{abstract} We consider the influence of extra dimensions on
the force in Casimir pistons.
Suitable analytical expressions are provided for
the Casimir force in
 the range where the plate distance is small,
and that where it is large, compared to the size of the extra
dimensions.
We show that the Casimir force
tends to move the center plate toward the closer wall; this
result is true independently of the cross-section of the piston
and
the geometry or topology of the additional Kaluza--Klein
dimensions. The statement also remains true at finite
temperature.
In the limit
where one wall of the piston is moved to infinity, the result for
parallel plates is recovered.
If only one chamber is considered, a criterion for
the occurrence of Lukosz-type repulsion,
as opposed to the occurrence of renormalization ambiguities, is given; 
we comment on why
no repulsion has been noted in some previous cosmological
calculations that consider only two plates.
 \end{abstract}

\preprint{NSF-KITP-08-145}
\maketitle

\section{Introduction} \label{sec-intro}
In recent years Casimir
pistons have received an increasing amount of interest because
they   allow the unambiguous prediction of forces,  free of the
  divergences that often plague Casimir calculations.
In their modern form they were introduced by Cavalcanti
\cite{cava04-69-065015} in a
two-dimensional setting. Namely, in his paper a Casimir piston
consists of a rectangular box divided by a movable partition
into two compartments, $A$ and $B$, of dimensions $a\times b$
and $(L-a)\times b$, respectively. Imposing Dirichlet boundary
conditions, as $L\to \infty$ it is shown that the piston is
attracted to the nearest end of the box. Higher-dimensional
pistons have been considered with various boundary conditions
\cite{hert05-95-250402,hert07-76-045016,mara07-75-085019,
eder07-75-105012,eder08-78-025021,lim0812.0426,lim0808.0047}.
Hertzberg et al showed that in three dimensions for perfect
metallic boundary conditions the rectangular piston is attracted
to the closest base \cite{hert05-95-250402,hert07-76-045016}; for
pistons with rectangular cross sections and Dirichlet or
Neumann boundary conditions see also
\cite{eder07-75-105012,eder07-09-005}. The same conclusion was
reached in \cite{eder08-78-025021} for perfect magnetic conductor
(infinitely permeable)
boundary conditions in a rectangular piston of arbitrary
dimension. Finally, a unified treatment reached the same
conclusion for  a scalar field with periodic, Dirichlet,
or Neumann boundary conditions and an electromagnetic field
with perfect electric conductor or perfect magnetic conductor
boundary conditions
\cite{lim0808.0047}. However, with the judicious choice of a
perfectly conducting piston inside a closed cylinder of arbitrary
cross section with infinitely permeable walls,
or a Dirichlet piston with Neumann walls, etc.,  a repulsive
force is found \cite{li97-56-2155,zhai07-76-047704,lim0812.0426};
those results generalize the famous observation of Boyer
\cite{boy74-09-2078} for parallel plates of unlike nature.
This work  mostly considered pistons of rectangular
cross section, where closed answers can be obtained. An exception
is \cite{mara07-75-085019} where it was shown that in three
dimensions a piston of arbitrary cross section, with all surfaces
perfectly conducting,
 is attracted to the closest wall.

In the limit where the transversal dimensions as well as one of
the walls is sent to infinity the configuration of two parallel
plates is obtained. The scalar Casimir force between parallel
plates in
the presence of compactified extra dimensions has been used to
put restrictions on the size of the extra dimensions
\cite{popp04-582-1,chen06-643-311,fran07-76-015008,fran08-78-055014,
chen08-668-72,obou0810.1096};
for  more recent discussions of perfectly conducting parallel
plates affecting the five-component electromagnetic field in five
dimensions (reaching different conclusions)
see \cite{pasc08-38-581,eder08-12-035}.
The Casimir effect was also used to argue
against the possibility that vacuum energy plays the role of a
cosmological
constant responsible for the observed dark energy
\cite{dora06-08-010,maha06-641-6,peri08-77-107301}; cutoff scales
that could produce the needed dark energy led to the prediction
of repulsive Casimir forces at distances between the plates where
attraction is verified experimentally
\cite{mohi98-81-4549,lamo97-78-5,bres02-88-041804}.

In this article we consider three-dimensional pistons of
arbitrary cross section in the context of Kaluza--Klein models.
We
show that in a scalar field theory with Dirichlet or Neumann
boundary conditions the piston is attracted to the closest wall
and that this statement holds independently of the cross-section
of the piston,  of the geometry and topology
of the additional Kaluza--Klein dimensions (with one minor
caveat),  and of the temperature.
We use zeta function techniques
\cite{dowk76-13-3224,hawk77-55-133,eliz94b,eliz95b,byts96-266-1,kirs02b}
to find closed answers for the Casimir force and to show these
results. In the Appendix indications are made how to use path
sums (or images) and an ultraviolet cutoff  to reach the same
conclusions.
It should be noted that mathematically the Kaluza--Klein
dimensions and the transverse dimensions of the macroscopic
piston play very similar roles, only their respective magnitudes
relative to the plate separation being quantitatively
significant.

The correction to the force due to the additional dimensions is
exponentially damped as long as the distance between the plates
is large compared to the size of the extra dimensions. However,
if the distance is comparable to or smaller than the size of the
extra dimensions the standard Casimir force in the space of the
full dimension is found. The crossover between the two regimes is
complicated, and different representations clearly showing the
different behavior are provided. The expansion in terms of the
ratio of distance between the plates over size of other
dimensions clearly shows how the geometry of the cross section
and of the Kaluza--Klein dimensions enters.

Furthermore, it is shown that a Lukosz-type repulsive force
\cite{luko71-56-109} can  appear only in a naive calculation
where
just one chamber of the piston is taken into account; for
related remarks see \cite{hert05-95-250402,hert07-76-045016}.
Moreover, an unambiguous prediction in such a case is
 possible only if a particular geometric
invariant of the extra dimensions vanishes.
A crucial point  is that this issue arises in the presence of
Kaluza--Klein dimensions even when the large transverse
dimensions are infinite; indeed, it is even more cogent there
than for a macroscopic box.
The present paper developed  from  a commentary
\cite{full09-671-179}
on these points as they arose in the papers of Cheng
\cite{chen06-643-311,chen08-668-72},
and it fulfills our pledge there to publish the details of our
calculations.
There is some overlap with an article by Teo \cite{teo0812.4641}
that appeared on the archive in the meantime and gives special
attention to the finite temperature theory.
The recent cosmological papers
\cite{popp04-582-1,fran07-76-015008,fran08-78-055014,
pasc08-38-581} do not explicitly consider the outer chamber of a
piston, but nevertheless they do not report a repulsive force; we
investigate the reason for that apparent discrepancy.

The article is organized as follows. We start by considering
parallel plates in the presence of extra dimensions. We first
find the Casimir force resulting from the space between the
plates only; the features just outlined are derived. We then add
the contributions from the exterior space to obtain the
generically attractive force between the plates. Section
\ref{cross-section} generalizes the results to an arbitrary
cross-section of the piston. Results for plate distances large,
respectively small, compared to the size of the extra dimensions
are given. For the case of a torus as Kaluza--Klein manifold more
explicit results are provided. In the Conclusions we summarize
the main results of the article and add pertinent remarks about
the finite-temperature case.

\section{Parallel plates in Kaluza--Klein models: Contributions
from between the plates} \label{naive}
Let $M=\reals^3\times N$.
We want to consider a piston geometry that lives in the
three-dimensional real space, and where there are additional
dimensions
described by the smooth Riemannian manifold $N$ of dimension $d$.
We realize the parallel plates as obtained from a piston geometry
with appropriate dimensions sent to infinity: Considering one
chamber of the piston with two dimensions already sent to
infinity, only two parallel plates a distance $D$ apart remain.
The correct answer for the parallel plates is obtained by adding
up answers for $D=a$ and $D=L-a$ sending $L\to\infty$. In this
section we deliberately consider only one chamber to highlight
the
serious flaws of this procedure.

We consider a scalar field model with Dirichlet boundary
conditions on the plates. The relevant eigenvalue spectrum of the
Laplacian on $M$ then is
$$\omega^2 = k_1^2+k_2^2+\left(
\frac{n\pi} {D}\right)^2 + \lambda_i^2,$$
where $n$ and $i$ are
positive integers, $k_1^2+k_2^2$ comes from the two free
transversal dimensions in $\reals^3$, $(n\pi/D)^2$ results from
the Dirichlet plates ($D=a$ for the left chamber and $D=L-a$ for
the right chamber), and $\lambda_i^2$ are the eigenfrequencies in
the additional dimensions,
\beq -\Delta_N \varphi_i= \lambda_i^2
\varphi_i.\label{Neigen}\eeq

The zeta function (density)
associated with this spectrum is
\beq \zeta (s) = \frac 1
{4\pi^2} \intl_{-\infty}^\infty dk_1 \intl _{-\infty} ^\infty
dk_2 \sum_{n=1}^\infty \sumi \left[ k_1^2 + k_2^2 +\left(
\frac{n\pi} D\right)^2 + \lambda_i^2\right]^{-s} .\label{pp1}\eeq
Performing the $k_1$ and $k_2$ integration we find
\beq \zeta (s)
= \frac 1 {4\pi (s-1)} \sumne \sumi \left[ \left( \frac{n\pi} D
\right)^2 + \lambda_i^2 \right]^{-s+1} . \label{pp2}\eeq
In order
to write down the necessary analytical continuation of this
expression, as is standard, a resummation of the $n$-summation is
applied. In that process, the zero modes $\lambda_j=0$ need
separate treatment. Letting $g_0$ be the degeneracy of the zero
eigenstate and assuming that $\lambda_i \geq 0$ we can write
\beq
\zeta (s) &=& \frac{g_0} {4\pi (s-1)} \left( \frac D \pi
\right)^{2s-2} \zeta _R (2s-2) \label{pp3}\\ & & + \frac 1 {4\pi
(s-1)} \sumne \sumip \left[ \left( \frac {n\pi} D \right)^2 +
\lambda_i^2\right] ^{-s+1} ,\nn\eeq
where the prime at the
$i$-summation indicates that the modes with $\lambda_j = 0 $ are
to be omitted from the summation. We rewrite the $n$-summation as
$$\sum_{n=1}^\infty = \frac 1 2 \left( \sum_{n=-\infty} ^\infty
-\,\, (n=0)\right),$$
and the $n=0$ term causes the occurrence of
the zeta function related to the eigenvalue problem
(\ref{Neigen}) on $N$,
 \beq \zeta_N (s) = \sumip \lambda_i^{-2s}.\label{pp4}\eeq
Using a Mellin transform this
allows the rewriting of (\ref{pp3}) as
\beq \zeta (s) &=& \frac
{g_0}{4\pi (s-1)} \left( \frac D \pi \right) ^{2s-2} \zeta _R
(2s-2) - \frac 1 {8\pi (s-1) } \zeta _N (s-1) \nn\\ & &+ \frac 1
{8\pi \Gamma (s) } \intl_0^\infty dt\, t^{s-2} \sumnm \sumip
e^{-\left[ \left( \frac{n\pi} D \right)^2 + \lambda_i^2\right]t}
. \label{pp5}\eeq
The last term is suitably manipulated employing
for $\alpha \in \reals_+$ \cite{hill62b}
 \beq \sumnm e^{-\alpha
n^2} = \sqrt{\frac \pi \alpha} \sumnm e^{-\frac{\pi^2 n^2} \alpha
} .\label{pp6}\eeq
As a result, for $n\neq 0$ we encounter the
integral representation of modified Bessel functions 
\cite{grad65b}
$$K_\nu (zx) = \frac{z^\nu} 2 \intl _0^\infty \exp \left\{ -\frac
x 2 \left( t+\frac {z^2} t \right) \right\} t^{-\nu -1} dt.$$
This allows us to obtain \beq \zeta (s) &=& \frac {g_0}{4\pi
(s-1)} \left( \frac D \pi \right) ^{2s-2} \zeta _R (2s-2) - \frac
1 {8\pi (s-1) } \zeta _N (s-1)
+ \frac{ D \Gamma \left( s-\frac 3 2 \right)} {8 \pi^{3/2} \Gamma
  (s) } \zeta _N \left( s- \frac 3 2 \right) \nn\\ & &+
  \frac{D^{s-1/2}}{2 \pi^{3/2} \Gamma (s)} \sumne \sumip \left(
  \frac{n^2} {\lambda_i^2}\right) ^{\frac 1 2 \left( s-\frac 3 2
  \right)} K_{\frac 3 2 - s} \left(
  2Dn\lambda_i\right).\label{pp7}\eeq

In order to find the
  Casimir energy, and then the force, for this setting, we need
  to evaluate this expression about the value $s=-1/2$. Whereas
  the first and last term are well defined at $s=-1/2$, and thus
  $s=-1/2$ can simply be substituted there, more care is needed
  for the second and third term. From general theory, see e.g.
  \cite{gilk95b,seel68-10-288,kirs02b}, it is known that $\zeta_N
  (s-1)$ will have a pole at $s=-1/2$ and that $\zeta_N (s-3/2)$
  will not vanish at $s=-1/2$. With $s=-1/2 +\epsilon$ and
  expanding about $\epsilon =0$ we therefore write
\beq \zeta_N
  (s-1) &=& \zeta_N \left( - \frac 3 2 + \epsilon \right) = \frac
  1 \epsilon \mbox{ Res } \zeta_N \left( - \frac 3 2 \right) +
  \mbox{ FP } \zeta_N \left( - \frac 3 2 \right) + {\cal O}
  (\epsilon ) , \nn\\
\zeta _N \left( s- \frac 3 2 \right) &=&
  \zeta _N (-2 + \epsilon ) = \zeta _N (-2) + \epsilon \zeta_N '
  (-2) + {\cal O} (\epsilon ^2) , \nn\eeq
and so
\beq \frac 1
  {s-1} \zeta_N (s-1) &=& - \frac 2 3 \left( \left[ \frac 1
  \epsilon + \frac 2 3 \right] \mbox{ Res } \zeta_N \left( -
  \frac 3 2 \right) + \mbox{ FP } \left( - \frac 3 2 \right)
  \right) + {\cal O} (\epsilon ) , \nn\\
\frac{ \Gamma \left( s-
  \frac 3 2 \right)} {\Gamma (s)} \zeta_N \left( s- \frac 3 2
  \right) &=& - \frac 1 {4 \sqrt \pi} \left( \zeta_N (-2) \left[
  \frac 1 \epsilon - \frac 1 2 + 2 \ln 2 \right] + \zeta_N ' (-2)
  \right) + {\cal O} (\epsilon ).\nn\eeq
This allows us to obtain
  \beq \zeta \left( - \frac 1 2 + \epsilon \right) &=& -
  \frac{\pi^2 g_0} {720} \frac 1 {D^3} + \frac 1 {12 \pi} \left(
  \mbox{ Res }\zeta_N \left( - \frac 32 \right) \left[ \frac 1
  \ep + \frac 2 3 \right] + \mbox{ FP } \zeta_N \left( - \frac 3
  2 \right) \right) \nn\\
& &- \frac D {32 \pi^2} \left( \zeta_N (-2) \left[ \frac 1 \ep
- \frac 1 2 + 2 \ln 2 \right] + \zeta_N ' (-2) \right) \label{pp8} \\
& &- \frac 1 {4\pi^2 D} \sumne \sumip \frac{\lis} {n^2}
K_2 (2Dn\lambda_i ).\nn\eeq
The resulting force {\it from one chamber} therefore reads
\beq F = - \frac 1 2 \frac \partial {\partial D} \zeta
\left(- \frac 1 2 + \ep \right) &=&
- \frac{\pi^2 g_0} {480 D^4} + \frac 1 {64 \pi^2}
\left( \zeta_N (-2) \left[ \frac 1 \ep - \frac 1 2 + 2 \ln 2 \right]+
 \zeta_N ' (-2) \right) \nn\\
& &+ \frac 1 {8\pi^2} \sumne \sumip \frac{\lis} {n^2}
 \frac \partial {\partial D} \frac 1 D K_2 (2Dn\lambda_i).\label{pp9}\eeq

Because of the pole at $\epsilon=0$ in \eqn{pp9},
for $\zeta_N (-2) \neq 0$ the zeta method leaves a
finite renormalization ambiguity proportional to $\zeta_N (-2)$.
 From a theoretical point of view no prediction about the sign of
the force can be made.
 In case $\zeta_N (-2) =0$ the force
appears to be finite.
(In the setting of an ultraviolet cutoff (see the Appendix) there
are additional divergences, but we shall not discuss them here.)
This can happen under certain restrictions
on the geometry of the manifold $N$. If we define $K_N(t)$ to be
the heat kernel associated with the eigenvalue problem
(\ref{Neigen}), its asymptotic expansion reads
\beq K_N (t) =
\sum_{i=1}^\infty e^{-\lambda_i^2 t} \sim
\sum_{l=0,1/2,1,...}^\infty b_l \,\,t^{l-\frac d 2}.
\label{heatN}\eeq
The heat-kernel coefficients are determined in
terms of geometric tensors of $N$ and its boundary, if present;
for a collection of known results see
\cite{gilk95b,gilk04b,kirs02b,vass03-388-279}. Using
$$\zeta _N (-2) = 2 b_{\frac {d+4} 2} =0$$
\cite{seel68-10-288},
we thus have a geometric condition on when
the force becomes
finite. When this vanishing occurs,
 $\zeta_N ' (-2) <0$
indicates that the force is definitely negative (attractive).
 If $\zeta_N ' (-2) >0$,
 asymptotically for $D\gg 1$ the force seems to be positive and
turns
negative at some critical distance $D_{crit}$. We come back to
this discussion in Section \ref{torus} when $N$ is chosen to be a
torus and where indeed $\zeta_N (-2) =0$.

\section{Parallel plates in Kaluza--Klein models: adding exterior
contributions}
\label{correct}

Let us now take into account the second chamber of the piston.
(In the present context that merely means adding a third plate at
a large distance.)
Denoting the plate separations by $a$ and $L-a$ and the
associated zeta functions by $\zeta_a (s)$ and $\zeta_{L-a} (s)$,
from (\ref{pp8}) one has immediately
\beq \zeta_a \left( - \frac
1 2 + \ep \right) + \zeta_{L-a} \left( - \frac 1 2 + \ep \right)
&=& - \frac{\pi^2 g_0} {720} \frac 1 {a^3} - \frac{ \pi^2 g_0}
{720} \frac 1 {(L-a)^3}\nn\\
 & & + \frac 1 {6\pi} \left( \mbox{ Res } \zeta_N \left( - \frac 3 2 \right)
\left[ \frac 1 \ep + \frac 2 3 \right]
 + \mbox{ FP } \zeta_N \left( - \frac 3 2 \right) \right) \nn\\
& &- \frac L {32 \pi^2} \left( \zeta_N (-2)
\left[ \frac 1 \ep - \frac 1 2 + 2 \ln 2 \right] +
\zeta_N ' (-2)\right)\label{pp10}\\
& &-\frac 1 {4\pi^2 a} \sumne \sumip \frac{\lis}{n^2}
 K_2 \left( 2an\lambda_i \right) \nn\\
& &-\frac 1 {4\pi^2 (L- a)} \sumne \sumip \frac{\lis}{n^2}
 K_2 \left( 2(L-a)n\lambda_i \right) .\nn\eeq
Despite the fact
that the Casimir \emph{energy} in general needs renormalization,
the
\emph{force} this time is always well-defined no matter what the
geometry or topology of $N$ looks like. In particular,
\beq F
&=& - \frac 1 2 \frac \partial {\partial a} \left[ \zeta_a \left(
- \frac 1 2 + \ep \right) + \zeta_{L-a} \left( - \frac 1 2 + \ep
\right)\right]\nn\\
&=& - \frac{\pi^2 g_0}{480 a^4} + \frac {\pi^2 g_0}
{ 480 (L-a)^4} \label{pp11}\\
& &+ \frac 1 {8\pi^2 }\sumne \sumip \frac{\lis} {n^2}
\frac \partial {\partial a} \frac 1 a K_2 (2an\lambda_i ) \nn\\
& &+ \frac 1 {8\pi^2 }\sumne \sumip \frac{\lis} {n^2}
 \frac \partial {\partial a} \frac 1 {L-a} K_2 (2(L-a)n\lambda_i ) .\nn\eeq
The force vanishes for $a=L/2$,
 is negative for $a<L/2$ and is positive for $a>L/2$;
 that is, the plate at $a$ is always attracted to the closer
wall.
As $L\to\infty$ the very simple result
\beq
F=-\frac{\pi^2g_0}{480 a^4} + \frac 1 {8\pi^2} \sumne \sumip
\frac{\lis} {n^2} \frac \partial {\partial a} \frac 1 a K_2
(2an\lambda_i)\label{pp12}\eeq
is obtained. The force is negative
independently of any details of the topology or geometry of the
extra dimensions (within the confine $\lambda_i \geq 0$).

\section{Pistons with finite cross section}
\label{cross-section}
In this section we  show that  a negative
force is  guaranteed whenever the boundary conditions on the
plates are both Dirichlet (or both Neumann), no matter what the
cross section
${\cal C}$ and the manifold $N$ are. Assume two parallel plates
 of some arbitrary shape within a cylinder of
that same
cross section, along with general Kaluza--Klein dimensions.
With
Dirichlet boundary conditions on the plates
(at separation $D$),
this gives rise to a
spectrum of the form
 \beq \omega^2=\left(\frac{n\pi} D\right)^2
+\mu_i^2.\label{specnofree}\eeq
The part $\mu_i^2$ comes from the
manifold $T={\cal C}\times N$. Proceeding exactly as before,
denoting
$$\zeta_T (s) = {\sumi} ' \mu_i^{-2s},$$
we obtain
\beq
\zeta (s) &=& \pi^{-2s} D^{2s} g_0 \zeta_R (2s) - \frac 1 2
\zeta_T (s) + \frac{ D \Gamma \left( s-\frac 1 2 \right) }{2\sqrt
\pi \Gamma (s) } \zeta_T \left( s-\frac 1 2 \right) \nn\\
& &+\frac{2 D^{s+1/2}}{\sqrt \pi \Gamma (s)} \sumne \sumip \left(
\frac{n^2}{\mu_i^2}\right)^{1/2 (s-1/2)} K_{1/2-s} \left(
2Dn\mu_i\right).\label{onechaarb}\eeq
The zeta function for the piston is the sum of two
such contributions
with $D$ replaced by $a$ and $L-a$ respectively.
 For the force this gives
\beq F&=& - \frac{\pi g_0} {24 a^2} + \frac{\pi g_0} {24 (L-a)^2}
\label{pistonarbcro}\\
&+& \frac 1 {2\pi} \sumne \sumip \frac{\mu_i} n
\frac \partial {\partial a} K_1 (2an\mu_i)\nn\\
&+& \frac 1 {2\pi} \sumne \sumip \frac{\mu_i} n
\frac \partial {\partial a} K_1 (2(L-a)n\mu_i).\nn\eeq
Again it is clearly seen that even in this generalized scenario
the
piston is attracted to the closest wall. In the limit
$L\to\infty$, the force reduces to
\beq F=-\frac{\pi g_0} {24 a^2} +
\frac 1 {2\pi} \sumne \sumip \frac{\mu_i} n \frac \partial
{\partial a} K_1 (2an\mu_i),\label{finaloneD}\eeq
which again is
manifestly negative.

The essential difference between this calculation and that of
Sec.~\ref{correct} is that we have only one infinite dimension
instead of three; up to this point the transverse and the
Kaluza--Klein dimensions have played identical roles. If we had
carried out the analog of Sec.~\ref{naive}, we would have
encountered similar results.  In particular, the famous repulsive
force of Lukosz \cite{luko71-56-109}
corresponds to rectangular cross section and $N=\emptyset$.

Although the representation \eqn{pistonarbcro} is most suitable
for reading off the sign
of the force, it is not useful numerically unless the plate
separation is sufficiently large compared to other scales.
 An expression suitable for reading off the small-$a$ behavior
is found by following formally a procedure used for
large-mass expansions \cite{kirs93-48-2813}, where the role of
the mass is played by the large parameter $\pi /a$. We first
rewrite the zeta function associated with the spectrum
(\ref{specnofree}) as
\beq \zeta (s) = \frac 1 {\Gamma (s)}
\sumne \sumi \int\limits_0^\infty dt \,\, t^{s-1} e^{-\left[
\left( \frac{n\pi } D \right)^2 + \mu_i^2\right] t}
.\label{zetacro}\eeq
We note that the small-$D$ expansion follows
from the small-$t$ behavior of the heat-kernel,
 \beq K(t) = \sumi
e^{-\mu_i^2 t} \sim \sum_{l=0,1/2,1,...}^\infty a_l
\,\,t^{l-\frac{d+2} 2} ;\nn\eeq
note that the spectrum $\mu_i^2$
results from a second-order partial differential operator in
dimension $(d+2)$.

Substituting this expansion into (\ref{zetacro}), asymptotically
as $D\to 0$ we find
\beq \zeta (s) = \frac 1 {\Gamma (s)}
\sum_{l=0,1/2,1,...}^\infty a_l \,\,\Gamma \left( s+l-\frac{d+2}
2 \right) \left( \frac D \pi \right)^{2s+2l-d-2} \zeta_R
(2s+2l-d-2).\nn\eeq
We evaluate this about $s=-1/2$ using well
known properties of the $\Gamma$-function and of the Riemann zeta
function \cite{grad65b}. The answers for $d$ even and odd look
slightly different; we denote by ${\sum_l}^d$ the summation
over $l=0,1,...,(d+2)/2,l>(d+4)/2$ for $d$ even, but over
$l=1/2,3/2,...,(d+2)/2,l>(d+4)/2$ for $d$ odd.
Furthermore, $\lfloor x\rfloor$
denotes the greatest integer not larger than $x$.
With $s=-1/2 +
\epsilon$, we get
\beq \zeta \left( - \frac 1 2 + \epsilon
\right) &=& - \frac 1 {2\sqrt \pi} {\sum_l}^d a_l \,\,\Gamma
\left( l-\frac{d+3} 2 \right) \left( \frac D \pi \right)
^{2l-d-3} \zeta_R (2l-d-3) \nn\\
&+& \sum_{j=1}^{\left\lfloor\frac{d+3} 2 \right\rfloor}
a_{\frac{d+3} 2 -j}
 \,\,\frac{(-1)^{j+1}}{j! \sqrt \pi}
\left( \frac \pi D \right)^{2j} \zeta_R ' (-2j)\nn\\
&+& a_{\frac{d+3} 2 } \left( \frac 1 {4 \sqrt \pi \epsilon}
+ \frac{\ln (4D) -1} {2\sqrt \pi } \right) \nn\\
&+&a_{\frac{d+4} 2 } \left( - \frac D {4\pi \epsilon} +
\frac D {2\pi} \left( 1-\gamma +
 \ln \left( \frac \pi D \right) \right) \right) +
{\cal O} (\epsilon ).\nn\eeq
This result is used to find the force from the left chamber. The
contribution to the force from the right chamber, where $D=L-a$
with $L\to\infty$, follows easily to be
 \beq F_2 &=& \frac 1
{8\pi \epsilon } \zeta_T (-1) + \frac 1 {8\pi } \left( \zeta_T '
(-1) + \zeta_T (-1) \left[ - 1 + \ln 4\right] \right) \nn\\
 &=& -
\frac 1 {8\pi} a_{\frac{d+4} 2 } \left( \frac 1 \epsilon - 1 +
\ln 4\right) + \frac 1 {8\pi} \zeta_T ' (-1).\nn\eeq
Here we used
$\zeta_T (-1) = - a_{(d+4)/2}$. As before, when the forces are
added, the  terms singular as $\epsilon \to 0$ cancel
and, asymptotically as $a\to 0$, the unambiguous answer for the force is found,
\beq F &=&
\frac 1 {2\pi ^{3/2}} {\sum_l}^d a_l \,\,\Gamma \left(
l-\frac{d+1} 2 \right) \left( \frac a \pi \right) ^{2l-d-4}
\zeta_R (2l-d-3) \nn\\
&+& \frac 1 {\pi ^{3/2}} \sum_{j=1}^{\left\lfloor \frac{d+3} 2
\right\rfloor}
a_{\frac{d+3} 2 -j} \,\,\frac{(-1)^{j+1}}{(j-1)!} \left( \frac
\pi a \right)^{2j+1} \zeta_R ' (-2j) \label{mainforce}\\
& -& \frac 1 {4 \sqrt
\pi a } a_{\frac{d+3} 2 } + \frac 1 {8\pi} \zeta_T ' (-1) -
\frac 1 {4\pi} a_{\frac{d+4} 2 } \left( \ln \left(\frac {2\pi} a
\right) - \gamma - \frac 1 2 \right) .\nn\eeq
Explicit expressions for given cross sections ${\cal C}$ and
Kaluza--Klein manifolds $N$ are easily obtained from known
expressions for the heat-kernel coefficients
\cite{kirs02b,gilk95b,vass03-388-279}.

In particular, if there are no additional dimensions,
$N=\emptyset$,
and  Neumann boundary conditions are imposed on the
cylinder walls, then
\beq a_0 = (4\pi)^{-1} \mbox{vol} ({\cal C}),
\quad \quad a_{1/2} = \frac 1 4 (4\pi)^{-1/2}
\mbox{vol}(\partial {\cal C}), \nn\eeq
and the first few terms of
the expansion reproduce the results in
\cite{hert05-95-250402,hert07-76-045016}.

If the manifold $N$ is nonempty and has no boundary, this time imposing 
Dirichlet boundary conditions on the cylinder walls, one easily
finds
\beq a_0 = (4\pi )^{-\frac{d+2} 2} \mbox{vol} ({\cal C})
\mbox{vol} (N), \quad \quad a_{1/2} = -\frac 1 4
(4\pi)^{-\frac{d+1} 2} \mbox{vol}(\partial {\cal C})
\mbox{vol}(N), \nn\eeq
and the leading two terms in the
asymptotic expansion of the force read
\beq F&\sim & 2^{-d-3}
\pi^{\frac{d+3} 2} \mbox{vol} ({\cal C}) \mbox{vol} (N) a^{-d-4}
\left\{ \begin{array}{ll} \Gamma \left( -\frac{d+1} 2 \right)
\zeta_R (-d-3) & d \mbox{ even} \\ \frac{2 (-1)^{\frac{d+1} 2} }
{\left( \frac{d+1} 2 \right)!} \zeta_R ' (-d-3) & d \mbox{ odd}
\end{array} \right. \nn\\ & &- 2^{-d-3} \pi^{\frac{d+2} 2}
\mbox{vol} (\partial {\cal C}) \mbox{vol}(N) a^{-d-3} \left\{
\begin{array}{ll} \frac{(-1)^{d/2}} {(d/2)!} \zeta_R ' (-d-2) & d
\mbox{ even} \\ \frac 1 2 \Gamma \left( - \frac d 2 \right)
\zeta_R (-d-2) & d \mbox{ odd} \end{array}
\right.\label{smalla}\eeq
Higher orders would involve the
extrinsic curvature of $\partial {\cal C}$ and  the curvature
of $N$. It is clearly seen that as soon as the plate separation
gets small compared to the sizes of other dimensions, the Casimir
force between the plates is significantly modified, revealing
information about the volume, and at higher
order the curvature,
of the Kaluza--Klein dimensions.

\section{Torus as Kaluza--Klein manifold}
\label{torus}

The series over the Bessel functions in eqs.~(\ref{pp9}),
(\ref{pp12}) and (\ref{finaloneD}) are numerically suitable as
long as the argument of the Bessel function grows sufficiently
fast with the eigenvalues. If that is the case, taking into
account only a few eigenvalues will be enough as contributions
are exponentially damped. However, in order to analyze how the
Casimir force behaves when the distance between the plates is
smaller than the size of the extra dimensions,  a
different procedure is necessary, as we have seen, and in general
only asymptotic
answers can be obtained.

For the case where $N=T^d$ it is possible to obtain closed
answers
that allow  consideration of several limits exactly. For
simplicity let
us assume an equilateral torus of radius $R$,
with the two macroscopic transverse dimensions effectively
infinite. The relevant
eigenvalue spectrum then reads
\beq \omega^2 = k_1 ^2+k_2^2 +
\left( \frac{n\pi} D \right)^2 + \frac 1 {R^2} \sum_{i=1}^d
n_i^2.\label{pp13}\eeq
For reasons explained above, the
previously obtained representations involving the Bessel
functions cannot be used easily to analyze the range where $D\ll
R$. If only the asymptotic behavior as $D\to 0$ is wanted, the use of 
\eqn{mainforce} is sufficient. For the torus, however, it is possible to recover 
also the exponentially damped contributions as $D\to 0$. In fact,
all technical tools have been provided to find closed
expressions in that regime. In particular it is again the
resummation (\ref{pp6}) that is relevant, but it should be
applied to the toroidal dimensions and not to the dimension in
which Dirichlet conditions are applied.
Applying resummation to
all $d$-sums originating from the torus, the result for the left
chamber corresponding to eq.~(\ref{pp7}) reads
\beq \zeta (s) &=&
\frac{\pi^{\frac{3d} 2 -2s+1} \Gamma \left( s -\frac d 2
-1\right)}{4\Gamma (s)} \left( \frac R D \right)^d D^{2s-2}
\zeta_R (2s-d-2) \label{pp14}\\
& &+ \frac{\pi^{\frac d 2 -1}} {2\Gamma (s)} \left(\frac R D \right)^{d/2}
(R\,\,D)^{s-1} \nn\\
& &\times \sumne \sumtop \left( \frac{n^2} {n_1^2+...+n_d^2}
\right)^{\frac 1 2 \left( 1-s+\frac d 2 \right)}
K_{1+\frac d 2 -s} \left( \frac{2\pi^2 R} D n
\sqrt{ n_1^2+...+n_d^2}\right) .\nn\eeq
For the right chamber we are mostly interested in the limit
$D\to\infty$, and so eq.~(\ref{pp7}) is the appropriate form.
Because $$\lambda_i^2 = \frac 1 {R^2} \sum_{j=1}^d n_j^2\,, $$
the
zeta function $\zeta_N (s)$ turns out to be the Epstein function
\cite{epst03-56-615,epst07-63-205} \beq Z_d (s;R) = R^{2s}
\sumtop (n_1^2+...+n_d^2)^{-s} . \label{pp15}\eeq Its analytical
continuation is very well understood
\cite{ambj83-147-1,eliz95b,eliz94-35-6100,kirs91-24-3281,kirs93-26-2421}
and we get \beq \zeta (s) &=& \frac{\pi^{1-2s} D^{2s-2}}{4 (s-1)}
\zeta_R (2s-2) \label{pp16}\\
& &-\frac 1 {8\pi (s-1)} Z_d (s-1; R) + \frac D {8\pi^{3/2}}
\frac {\Gamma \left( s- \frac 3 2 \right)}{\Gamma (s)}
Z_d \left( s- \frac 3 2 ; R\right) \nn\\
& &+ \frac{D^{s-1/2} R^{s-3/2}}{2\pi^{3/2} \Gamma (s)}
\sumne \sumtop \left( \frac{n^2} {n_1^2+...+n_d^2}
\right)^{\frac 1 2 \left( s- \frac 3 2 \right)}
 K_{\frac 3 2 -s } \left( \frac{2Dn} R \sqrt{ n_1^2+...+n_d^2}\right).
\nn\eeq
Using this representation (\ref{pp16}) for both chambers,
and noting $g_0=1$,
 we get immediately from eq.~(\ref{pp12}) the force
\beq F &=& - \frac{\pi^4}{480 a^4} \label{pp17}\\
& &+\frac 1 {8\pi^2 R^2} \sumne \sumtop \frac{n_1^2+...+n_d^2}{n^2}
 \frac \partial {\partial a} \frac 1 a K_2
\left( \frac{ 2an } R \sqrt{ n_1^2+...+n_d^2}\right).\nn\eeq
This representation is particularly suitable for $R<a$ because
the contributions from $K_2$ are exponentially damped. It shows that
 as long as the size of the extra dimensions is small compared to
the separation of the plates, the correction to the well-known
Casimir
 force between parallel plates is very small.

The above representation is not suitable for the range with plate
separation smaller than $R$, because $K_2 (z) \sim 2/z^2$ as
$z\to 0$. As we will see, the leading contribution as $a\to 0$
will then come from the series. A better suited representation is
obtained by rewriting eq.~(\ref{pp17}) using the fact that eqs.\
(\ref{pp14}) and (\ref{pp16}) equal each other. This first shows
\beq & &\frac{D^{s-1/2} R^{s-3/2}}{2 \pi^{3/2} \Gamma (s)} \sumne
\sumtop \left( \frac{n^2}{\spt}\right) ^{\frac 1 2
\left( s- \frac 3 2 \right)} K_{\frac 3 2 -s } \left( \frac {2Dn} R
\sqrt{\spt}\right) \nn\\
&=& \frac{\pi^{\frac{3d} 2 -2s +1} \Gamma
\left( s-\frac d 2 -1\right)}{4\Gamma (s)}
\left( \frac R D \right)^d D^{2s-2} \zeta_R (2s-d-2) \nn\\
&+&\frac{\pi^{\frac d 2 -1}}{2\Gamma (s)} \left( \frac R D \right)^{d/2}
 (R \,\, D)^{s-1} \sumne \sumtop \left( \frac{n^2} {\spt}
\right)^{\frac 1 2 \left( 1-s+\frac d 2 \right)}\nn \\
& &{}\times
K_{1+\frac d 2 -s} \left( \frac{ 2\pi^2 R} D n \sqrt{\spt}\right) \nn\\
&-& \frac{\pi^{1-2s} D^{2s-2}}{4 (s-1)} \zeta_R (2s-2) +
 \frac 1 {8\pi (s-1)} Z_d (s-1; R) - \frac D {8\pi^{3/2}}
\frac{\Gamma \left( s-\frac 3 2 \right)} {\Gamma (s)}
Z_d \left( s-\frac 3 2 ; R\right).\nn\eeq
Analytically continuing this to $s=-1/2$, one obtains
\beq & &\frac 1 {8\pi^2 R^2} \sumne \sumtop \frac{ \spt} {n^2}
 \frac \partial {\partial a} \frac 1 a K_2 \left( \frac{2an} R
 \sqrt{ \spt}\right) =\nn\\
&-& (d+3) \frac{ \pi^{\frac 3 2 (d+1)}} {16} \left( \frac R a
\right)^d a^{-4} \left\{ \begin{array}{lr}
\Gamma \left( - \frac{d+3} 2 \right) \zeta_R (-d-3), & d \mbox{ even}\\
\frac{2 (-1)^{\frac{d-1} 2 } }{\left( \frac{d+3} 2 \right)!}
 \zeta_R ' (-d-3), & d \mbox{ odd}\end{array} \right.\nn\\
& &+\frac{\pi^2}{480 a^4} - \frac 1 {64 \pi^2} Z_d '
 \left(-2; R \right) \nn\\
& &+ \frac{ \pi^{\frac {d-3} 2 } R^{\frac {d-3} 2} }
 8 \sumne \sumtop \left( \frac{n^2}{n_1^2+...+n_d^2}\right)^{\frac {3+d} 4}
\frac \partial {\partial a} a^{-\frac{3+d} 2 } K_{\frac{d+3} 2 }
\left( \frac{2 \pi^2 R} a n \sqrt{ n_1^2+...+n_d^2} \right).\nonumber\eeq
Using this in eq.~(\ref{pp17}) then gives the force in the form
\beq F &=&  \frac{ \pi^{\frac 3 2 (d+1)}} {8}
\left( \frac R a \right)^d a^{-4} \left\{ \begin{array}{lr}
\Gamma \left( - \frac{d+1} 2 \right) \zeta_R (-d-3), & d \mbox{ even}\\
\frac{2 (-1)^{\frac{d+1} 2 } }{\left( \frac{d+1} 2 \right)!}
 \zeta_R ' (-d-3), & d \mbox{ odd}\end{array} \right.
\nn\\
& &- \frac 1 {64 \pi^2} Z_d ' \left(-2; R \right)
\label{nearforce}\\
& &+ \frac{ \pi^{\frac {d-3} 2 } R^{\frac {d-3} 2} }
 8 \sumne \sumtop \left( \frac{n^2}{n_1^2+...+n_d^2}\right)^{\frac {3+d} 4}
\frac \partial {\partial a} a^{-\frac{3+d} 2 } K_{\frac{d+3} 2 }
\left( \frac{2 \pi^2 R} a n \sqrt{ n_1^2+...+n_d^2} \right).\nonumber\eeq
This result shows that if $a \ll R$, the compactness of the extra
 dimensions can be ignored, and the force is the standard Casimir force,
 but in the space of the full dimension, namely of dimension $3+d$.
 The first term agrees with the general result (\ref{smalla}) when
 specialized to $N=T^d$.

For this example it is clear that $\zeta_N (-2) =Z_d (-2;R) =0$,
because the torus is a flat manifold without boundary.
Therefore the force resulting from one chamber only, as given in
(\ref{pp9}), is finite. Using the reflection formula for the
Epstein function \cite{ambj83-147-1}
it is obtained as
\beq F &=& - \frac{\pi^2} {480 a^4}
+ \frac{\Gamma \left( \frac d 2 +2\right)}
{32 \pi^{6+d/2} R^4} Z_d \left(\frac d 2 +2;1\right)
\label{naiveforce}\\
& &+\frac 1 {8\pi^2 R^2} \sumne \sumdp \frac{n_1^2+...+n_d^2}{n^2}
 \frac \partial {\partial a}
\frac 1 a K_2 \left( \frac{2an} R \sqrt{n_1^2+...+
n_d^2}\right),\nn\eeq
For $a \gg R$ this force, obtained by neglecting the second
chamber, is positive
and asymptotically constant.
 Considering $N$ to be a rectangle with Dirichlet or Neumann
boundary conditions leads to the same result \cite{chen06-643-311}.
Whatever may be the case for a macroscopic conducting box
(where there may not be an external piston shaft), in a
Kaluza--Klein cosmology the extra dimensions are indisputably
present outside the parallel plates as well as inside.
Therefore, formula \eqn{naiveforce} surely must be rejected as
spurious.

However, the papers
\cite{popp04-582-1,fran07-76-015008,fran08-78-055014,pasc08-38-581}
did not take the outer chamber of the Kaluza--Klein
piston into account, but nevertheless they did not find a
repulsive force.
Closer examination (see, for example, eqs.\ (24) and (25) of
\cite{pasc08-38-581}, or p.~5 of \cite{fran07-76-015008}) shows
that all those authors have indeed
subtracted the term (linear in $a$) that
we here consider to be the piston correction.
Their reasoning is to subtract the Casimir energy
(caused by the small compact dimensions) that would
exist in the region
between the plates if the plates were not there.
In the Cavalcanti piston, the
analogous reasoning would be to make the piston shaft infinite in
both
directions, ignore the outer chambers, remove the plates, and
subtract the energy  in the inner chamber.
We believe that our analysis is more convincing.

\section{Conclusions} \label{concl}
In this article we have analyzed forces occurring in pistons of
arbitrary cross section in a cosmological Kaluza--Klein setting.
We have shown that irrespective of the details of the cross
section and of the geometry and topology of the Kaluza--Klein
manifold, the piston is always attracted to the closest wall.
This implies that parallel plates always attract no
matter what the properties of the additional dimensions are
(except for the physically mild restriction that the eigenvalues
$\lambda_i^2$ or $\mu_i^2$ all be nonnegative).
Repulsive forces between Dirichlet plates can  occur only in a
naive
calculation that takes into account only one of the chambers; see
the explanations at the ends of Secs.\ \ref{naive} and \ref{torus}.
Furthermore, we have
derived an asymptotic expansion of the force for small distances
between the piston and the wall,  eq.~(\ref{smalla}). It is
clearly seen how the geometries of the cross section and  the
Kaluza--Klein manifold enter the answer. In this limit the
plates notice the dimensions of all space and the force obtained
is the standard Casimir force in the space of the full dimension.

All results for the force remain valid if Neumann boundary
conditions on both plates instead of Dirichlet boundary
conditions
are considered, because the Neumann and Dirichlet spectrum differ
only by $a$-independent eigenvalues.

The attraction of the piston to the closest wall is further enhanced by
 finite-temperature contributions. Assuming a piston with
arbitrary cross section, the relevant finite temperature spectrum reads
\beq \omega ^2 = \left( \frac {2\pi j} \beta \right)^2 +
\left( \frac{n\pi} D \right)^2 + \mu_i^2, \quad \quad j\in\intgs ,\nn\eeq
where $\beta$ is the inverse temperature.
 The energy associated with the system is defined to be
\cite{dowk76-13-3224,hawk77-55-133}
\beq E = - \frac 1 2 \frac \partial {\partial \beta}
\left[ \zeta ' _ \beta (0) + \zeta _\beta (0) \ln \mu^2\right],\nn\eeq
 $\zeta_\beta (s)$ as the zeta function arising
from this spectrum and $\mu$ is a renormalization scale.
Using as is
standard a resummation of the Matsubara sum,
 the following form can be obtained \cite{dowk78-11-895}:
\beq E&=&\frac 1 2 {\mbox FP } \zeta \left( - \frac 1 2\right) +
\mbox{Res } \zeta \left( - \frac 1 2 \right) \ln
\left( \frac{\mu e} 2 \right) + \sumne \sumi \frac{\lambda_{i,n} }
{\left(e^{\beta \lambda_{i,n}} -1\right)},\nn\eeq
where $\lambda_{i,n}^2 = \left( \frac{n\pi} D \right)^2 + \mu_i^2$ and
$\zeta (s)$ is the zeta function analyzed in Section \ref{cross-section}.
The temperature contribution is a decreasing function of $D$
\cite{teo0812.4641}. Thus, when the contributions of the two
chambers are added,
 the finite-temperature part, just like the zero-temperature part
discussed previously,
 tends to move the piston toward the closer wall.


\appendix\section{The method of images}

As in Section \ref{torus}, consider
two infinite transverse dimensions;  one Dirichlet plate
separation, $D$;
and $d$ periodic Kaluza--Klein dimensions, all of circumference
$2\pi R$.
The system can easily be treated by the methods of \cite{rect}
(for example).

The free cylinder kernel (a certain Green function for the
Laplacian in $\reals^{d+4}$) is
\beq
T_0= C_{d+3}\,t
[t^2+\|\mathbf{x}-\mathbf{x}'\|^2]^{-(d+4)/2},  \qquad
C_{d+3}=
\frac{\Gamma\left(\frac{d+4}2\right)}{\pi^\frac{d+4}2}\,.
\eeq
Thus
\beq  -\,\frac12\,\frac {\partial T_0}{\partial t} &=&
\frac12  C_{d+3} \left\{
-[t^2+\|\mathbf{x}-\mathbf{x}'\|^2]^{-(d+4)/2}
+ (d+4) t^2
[t^2+\|\mathbf{x}-\mathbf{x}'\|^2]^{-(d+6)/2}\right\}.
\label{cylderiv}\eeq
 The corresponding Green function $T(t,\mathbf{x},\mathbf{x'})$ 
in a rectangular geometry 
is given exactly by a sum over all classical paths from 
$\mathbf{x}'$ to $\mathbf{x}$, or, equivalently, by 
a sum over $T_0$ displaced to appropriate ``image 
charges''\negthinspace.
In the limit $t\downarrow0$, formally the energy density is
\beq
T_{00} = -\,\frac12\,\left.\frac{\partial T}{\partial t}
\right|_{t=0}
=-\frac12 \sum_{\mbox{\small images}}
C_{d+3}
\|\mathbf{x}-\mathbf{x}'\|^{-(d+4)}.
\label{edens}\eeq
For a careful study of divergences one would maintain the last
factor in the form
$[t^2+\|\mathbf{x}-\mathbf{x}'\|^2]^{-(d+4)/2}$.
Let $B_d= - \frac12 C_{d+3}\,$.

Let $\mathbf{N} = (N_1,\ldots,N_d) \in \intgs^d$, and let $\kh$
be the
unit vector perpendicular to the plates in the physical space
(the $z$ direction).
Periodic boundary conditions are imposed by summing displacements
of formula \eqn{edens} over a periodic lattice, and
the lattice of Dirichlet images is a difference of two periodic
lattices:
\beq T_{00}&=&B_d \sum_{N_1=-\infty}^\infty\cdots\sum_{N_d=-\infty}^\infty
\sum_{N=-\infty}^\infty
\{ \|\mathbf{x}-(\mathbf{x}   +2D\kh N+2\pi R
\mathbf{N})\|^{-(d+4)}) \nn \\
&-& \|\mathbf{x}-(\mathbf{x} -2z\kh  -2D\kh N-2\pi R
\mathbf{N})\|^{-(d+4)} \}\nn\\
&=&B_d \sum_{N_1=-\infty}^\infty \cdots\sum_{N_d=-\infty}^\infty
\sum_{N=-\infty}^\infty
\{ \|2D\kh N  +2\pi R \mathbf{N}|^{-(d+4)} \nn \\
&-& \|2z\kh + 2D\kh N +2\pi R \mathbf{N}\|^{-(d+4)} \}.
\label{imagesum} \eeq
To get the energy per unit area one should integrate over $z$ and
over the periodic coordinates.
The latter amounts to multiplying by $(2\pi R)^d$.

We now sketch the process of discarding divergent terms
(which appear in the present ultraviolet-cutoff method as
negative powers of~$t$, but were automatically eliminated by the
zeta-function regularization).
 The term $N=0$, $\mathbf{N}=0$ is
the free vacuum energy.
The other periodic terms with $N=0$ (the analog of terms called
PV in \cite{rect}) are
(after the integration)
\beq
B_d (2\pi R)^{-4} D\sum_{\mathbf{N}\ne0} \|\mathbf{N}\|^{-(d+4)}.
\label{PV}\eeq
This expression (with $D=a$) will add to the corresponding term
from the
piston shaft (with $D=L-a$)
to give an energy per area independent of $a$, hence no
pressure.
Without the shaft, however, \eqn{PV} (with $D=a$) gives a
repulsive Lukosz
pressure independent of $a$.
The remaining terms in the periodic orbit sum (the first term in
the final version of \eqn{imagesum}) are
\beq
B_d (\pi R)^{d}2^{-4}D \sum_{N\ne 0} \sum_{\mathbf{N}} \|\pi
R \mathbf{N} +
D\kh N\|^{-(d+4)}.  \label{PDH} \eeq
These are the PD and PH terms; \eqn{PDH} (with $D=a$) is the main
Casimir energy.
With more effort it could be shown to yield the same forces as
found in Sec.~\ref{torus}.

In the terminology of \cite{rect} there are no VP, VD, or C terms
in this problem, because there are no
reflections in the periodic directions.
The last term in \eqn{imagesum} consists
of HP terms ($\mathbf{N}=0$) and HD terms.
The first of these formally give the energy
\beq
-\mbox{(constant)}\int_0^D dz \sum_{N=-\infty}^\infty
|z+DN|^{-(d+4)}
= -\mbox{(constant)} \int_{-\infty}^\infty |z|^{-(d+4)} dz.
\label{HP}\eeq
The integral (which would converge if we had kept $t>0$) is
independent
of $D$ and hence gives no force.  This is the surface energy of
the plates (renormalizes their masses).
Finally, the HD terms, before integration over $z$, are
\beq
& -& B_d (\pi R)^d 2^{-4} \sum_{\mathbf{N}\ne 0}
\sum_{N=-\infty}^\infty
\|(z+DN)\kh +
\pi R \mathbf{N}\|^{-(d+4)} \nn\\
& =& - B_d (\pi R)^d 2^{-4} \sum_{\mathbf{N}\ne 0}
\sum_{N=-\infty}^\infty
[(z+DN)^2 + (\pi R)^2\|\mathbf{N}\|^2 ]^{-(d+4)/2}.
\nn\eeq
Upon integration the $N$ sum again telescopes:
\beq
-B_d (\pi R)^d 2^{-4} \sum_{\mathbf{N}\ne 0}
\int_{-\infty}^\infty
[z^2+ (\pi R)^2 \|\mathbf{N}\|^2]^{-(d+4)/2}\, dz.
\label{HD}\eeq
Again this contribution is independent of $D$ (being a surface
effect,
albeit dependent on the geometry of the extra dimensions).
The HD energy does not contribute to the force, just as in the
original Cavalcanti piston (or single box).

\subsection*{Acknowledgments}
Our work is supported by National Science Foundation Grants
PHY-0757791 (Baylor) and PHY-0554849 (TAMU). 
Much of the work was done while 
S.A.F. enjoyed the hospitality and partial support of the 
Institute for Mathematics and Its Applications (NSF Grant 
DMS-0439734) and the Kavli Institute for 
Theoretical Physics (NSF Grant PHY-0551164).
We thank Gerald Cleaver, Richard Obousy, and Felipe da Rosa for 
discussions.


\end{document}